\documentclass[showpacs,prl]{revtex4}

\usepackage{graphicx}
\usepackage{dcolumn}
\usepackage{bm}

\input{tcilatex}

\begin{document}

\title{Generalized Poland-Scheraga model for supercoiled DNA}

\author{Thomas Garel, Henri Orland \\
Service de Physique Th\'{e}orique (URA 2306 du CNRS)\\
CEA-Saclay, 91191 Gif-sur-Yvette cedex, France \\
and \\
Edouard Yeramian \\
Unit\'e de Bioinformatique Structurale (URA 2185 du CNRS) \\
Institut Pasteur, 75724 Paris cedex 15, France
}

\begin{abstract}
The Poland-Scheraga (PS) model for the helix-coil transition of DNA
considers the statistical mechanics of the thermally induced binding
of two complementary strands of DNA.
In this paper, we show how to modify the PS model 
when a torque is applied to the extremities of DNA:
We propose a simple model for the energy of twisted DNA and compute the
entropy of a loop, subject to angular
constraints (supercoiling).
The denaturation curves are shifted towards lower or higher
temperatures depending on the sign of the torque, and the UV
absorption peaks are softened. The properties of supercoiled DNA 
can be deduced through the use of a numerical Legendre
transform. In the homogeneous case, we find
that for weak supercoiling, the phenomenological
quadratic law relating the torsional energy to the number of unpaired
bases is recovered. 
\bigskip


\end{abstract}
\pacs{87.14.Gg; 87.15.Cc; 82.39.Pj}
\maketitle
Natural DNA exists as a double helix bound state \cite{CW}. Upon
heating, the two complementary strands may separate. This thermal unbinding
transition is called DNA denaturation (see \cite{Pol_Scher2} and
references therein). It has been modeled in various ways, the most
prominent being the Poland-Scheraga (PS) model \cite{Pol_Scher1}. Even
though this model does not take into account spatial aspects of the denaturation
transition, it correctly treats sequence effects, and has been
numerically implemented in the program MELTSIM \cite{Meltsim}. It has also been shown
that a mechanically induced denaturation transition is possible, by
the combined application of a stretching force and an untwisting
torque (see \cite{Strick} and references therein).
In a brief review of 
the standard PS model, we emphasize the role of the loop exponent $c$
on the existence and nature of the denaturation transition.
We generalize the theory to include a torque which 
introduces a torsional enthalpy term, and results in a modified loop
exponent $c' < c$. Numerical simulations on biological sequences show
that the denaturation curves are shifted, while the peaks are smoother
than their zero torque counterparts. Extension to supercoiled or
undercoiled DNA , through the use of a numerical 
Legendre transform, yields denaturation isotherms for the same sequences.


We briefly review the Poland-Scheraga (PS) model for DNA melting, and
consider a double stranded (ds) DNA fragment, made of $N$
complementary base pairs, assuming that bases $(1)$ and $(N)$ on both
strands are paired. We denote by $Z(\alpha )$ the forward partition
function of the two strands, starting at base $(1)$ and ending at base
$(\alpha )$, with bases $(\alpha )$ being paired. This partition
function satisfies the recursion relation (Figure 1)

\begin{equation}
Z(\alpha +1)=e^{-\beta \varepsilon _{\alpha ,\alpha +1}}\ Z(\alpha )+\sigma
_{S}\sum_{\alpha ^{\prime }=1}^{\alpha -1}Z(\alpha ^{\prime })\mathcal{N}%
(\alpha ^{\prime };\alpha +1)  \label{y1}
\end{equation}%
where $\beta =1/k_{B}T$ is the inverse temperature, $\varepsilon _{\alpha
,\alpha +1}$ is the stacking energy of base pairs $(\alpha ,\alpha +1)$, and 
$\sigma _{S}$ is the bare loop formation (cooperativity)\ parameter (we
assume that $\sigma _{S}$ is base independent).

\begin{figure}[htbp]
\begin{center}
\includegraphics[scale=1.0]{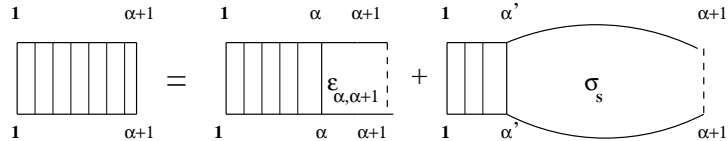}
\end{center}
\caption{Recursion relation for $Z(\protect\alpha +1)$ (eq.(\protect\ref{y1}%
)) in the PS model.}
\label{f1}
\end{figure}

The factor $\mathcal{N}(\alpha ^{\prime };\alpha +1)$ counts the number of
conformations of a pair of chains starting at base pair $(\alpha ^{\prime })$
and ending at base pair $(\alpha +1)$. It also represents the number of
conformations of a closed polymer of $2(\alpha-\alpha^{\prime})$ monomers,
which is asymptotically given by \cite{PGG}

\begin{equation}
\mathcal{N}(\alpha ^{\prime };\alpha +1)=\mu _{0}^{\alpha -\alpha ^{\prime
}}g(\alpha -\alpha ^{\prime })=\frac{\mu _{0}^{\alpha -\alpha ^{\prime }}}{%
(\alpha -\alpha ^{\prime })^{c}}  \label{asymp}
\end{equation}%
where $k_{B}\log \mu _{0}$ is the entropy per base pair (assumed to be
independent of the chemical nature of the pair), and $g(x)=\frac{1}{x^{c}}$
is the probability of return to the origin of a loop of length $2x$. The
exponent $c$ depends on the interaction of the loop with itself and with the
rest of the chain: It has been extensively discussed in the context of
homopolymeric DNA \cite{Pol_Scher1,Fisher,Ka_Mu_Pe}. If one neglects the
interaction with the rest of the chain, we have $c=3\nu $ (yielding $3/2$
for a Gaussian loop and $\approx 1.8$ for a self avoiding loop). Taking into
account the interaction with the rest of the chain is a difficult problem:
approximations and numerical calculations point toward a value $\approx 2.15$
for the full problem \cite{Dup,Orlandini}. 

The recursion relation (\ref{y1}) is supplemented by the boundary conditions
$Z(1) =1;\ Z(2) =e^{-\beta \varepsilon _{1,2}}Z(1)$.
This recursion relation can easily be solved analytically if one assumes
that all stacking energies are equal. One may for instance introduce a grand
canonical partition function $\mathcal{Z}(z)=\sum_{\alpha =1}^{\infty
}z^{\alpha }Z(\alpha )$ . We summarize the results of this homopolymeric
study: 
i) If $\ 2<c$, there is a first order (discontinuous) unbinding
transition.
ii) If \ $1<c<2~$, there is a second order (continuous) unbinding
transition, with a specific heat exponent $\alpha =\frac{2c-3}{c-1}$.
iii) If \ $c<1$ , the two strands are always bound and loops open in a
continuous way.

For non homogeneous sequences, the calculation cannot be done analytically.
However, the results pertaining to the existence of an unbinding transition
are expected to hold. In addition, in order to calculate the probability of
opening of a base pair, it is necessary to introduce forward and backward
partition functions \cite{Poland,Ga_Orl}. The forward partition
function $Z_{f}(\alpha )$ is nothing but $Z(\alpha )$ whereas the
backward partition function $ Z_{b}(\alpha )$ is the partition
function of the two strands, starting at base ($N$) and ending at base
($\alpha $), with base ($\alpha $) being paired. These points will be
discussed in detail in \cite{tobe}. 

We now generalize the Poland-Scheraga model to the case
where a torque is applied to the DNA fragment.


We again assume that base pairs (1) and (N) of the DNA fragment are kept
fixed and apply a weak torque $\Gamma $ on base pair (N). By
\textquotedblleft weak\textquotedblright , we mean that there are no
plectonemes on the chain: Experimentally, applying a force $F>0.5$ pN on a
DNA fragment of a few persistence lengths $l_{p}~(l_{p}\approx 150$ bp in
(ds) DNA) is enough to prevent the formation of plectonemes
\cite{Strick}.

\begin{figure}[htbp]
\begin{center}
\includegraphics[scale=0.7]{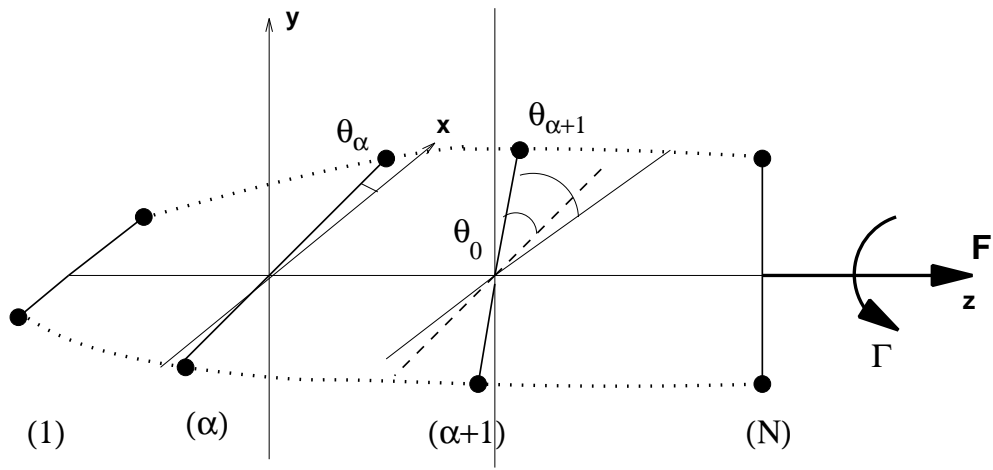}
\end{center}
\caption{Schematic representation of the force $F$ + torque $\Gamma$
experiment. Basepair ($\alpha$) has orientation ($\theta_{\alpha}$). The 
natural twist of the helix is ($\theta_0$).}
\label{f2}
\end{figure}

If the force $F$ \ is applied along the $z-$axis, the DNA fragment will be
aligned (on the average) along this direction. One may then assign, in 
the $xy$ plane, an angle $%
\theta _{\alpha }$ to a paired base pair $(\alpha )$, representing the angle
of this pair with the $x-$axis (fig. 2). We denote by $\theta _{0}$ the natural twist
angle of the DNA helix per base pair ($\theta _{0}=2\pi /10.4$ in radians),
and model the torsional energy between neighboring base pairs by 
\begin{equation}
\varepsilon _{0}(\alpha ,\alpha +1)=\frac{1}{2}\kappa _{0}(\theta _{\alpha
+1}-\theta _{\alpha }-\theta _{0})^{2}
\end{equation}%
where $\kappa _{0}$ is the elastic torsion constant of (ds) DNA.

Generalizing the PS model, one may define a partition function $Z(\alpha
,\theta _{\alpha })$ of the two strands, starting at base pair $(1)$ with
orientation $\theta _{1}$, and ending at base pair $(\alpha )$, with
orientation $\theta _{\alpha }$. This partition function satisfies 
\[
Z(\alpha +1,\theta _{\alpha +1})=\ e^{-\beta \varepsilon _{\alpha ,\alpha
+1}}\int_{-\infty }^{+\infty }{\frac{d\theta _{\alpha }}{M_{0}}}e^{-{\frac{%
\beta \kappa _{0}}{2}}(\theta _{\alpha +1}-\theta _{\alpha }-\theta
_{0})^{2}}\ Z(\alpha ,\theta _{\alpha }) 
\]%
\begin{equation}
+\sigma _{S}\sum_{\alpha ^{\prime }=1}^{\alpha -1}\int_{-\infty }^{+\infty }%
\frac{d\theta _{\alpha ^{\prime }}}{M_{1}}{e}^{-\beta E_{1}\bigl(\alpha
^{\prime },\theta _{\alpha ^{\prime }};\alpha +1,\theta _{\alpha +1}\bigr)}%
\mathcal{N}\bigl(\alpha ^{\prime },\theta _{\alpha ^{\prime }};\alpha
+1,\theta _{\alpha +1}\bigr)Z(\alpha ^{\prime },\theta _{\alpha ^{\prime }})
\label{z1}
\end{equation}%
where $\varepsilon _{\alpha ,\alpha +1}$ again denotes the stacking energy
of base pairs ($\alpha ,\alpha +1$) and ($M_{0},M_{1}$) are normalization
factors (see below).

In equation (\ref{z1}), the existence of the stretching force $F$ is
implicit. The functions $E_{1}\bigl(\alpha ^{\prime },\theta _{\alpha
^{\prime }};\alpha +1,\theta _{\alpha +1}\bigr)$ and $\mathcal{N}\bigl(%
\alpha ^{\prime },\theta _{\alpha ^{\prime }};\alpha +1,\theta _{\alpha +1}%
\bigr)$ represent respectively the torsional energy and the number of
conformations of a pair of chains starting at base pair ($\alpha ^{\prime }$%
) with orientation ($\theta _{\alpha ^{\prime }}$) and ending at base pair $%
(\alpha +1)$ with orientation $(\theta _{\alpha +1})$.

We first discuss $\mathcal{N}\bigl(%
\alpha ^{\prime },\theta _{\alpha ^{\prime }};\alpha +1,\theta _{\alpha +1}%
\bigr)$. With unconstrained orientations, equation (\ref{asymp})
emphasizes the importance of the loop exponent $c$, representing the
interaction of the loop with itself and with the rest of the chain. In
a forthcoming paper \cite{tobe}, we show that $c$ can also be modeled
through the introduction of a specific repulsive potential in an 
otherwise Gaussian loop. In a nutshell, the two strands partition
function can be factorized as a product of a center of mass partition function
$Z_{cm}(F)$ and a relative coordinate partition function
$$
Z_{\rho}=\int {\cal D}\vec {\rho}(s) \ e^{-\beta H_{\rho}}
$$
with
$\beta H_{\rho}=\frac{3}{4 a^2}\ \int ds \ {\vec {\dot
\rho}^2(s)}+D \int ds \ \frac{1}{\vert\vec
\rho(s)\vert^{\lambda}}  
$
, where $a \approx 50$ \AA $ \ $is the single strand (ss) DNA Kuhn
length. The effective chain described by $\vec \rho(s)$ has length
$2(\alpha-\alpha')$, and the probability of return to the origin is
given by 
\begin{equation}
\label{probapot}
g_{D}(\alpha -\alpha')={ <0\vert e^{-2(\alpha -\alpha')(\beta
h_{D})} \vert 0> \over
\int dr <r \vert e^{-2(\alpha -\alpha')(\beta
h_{D})} \vert 0>}
\end{equation}
with $\beta h_{D}=-\frac{a^2}{3}\nabla_{\rho}^2 +\frac{D}{\rho^{\lambda}}$.

This model will mimick the original problem, if one has
$g_{D}(\alpha -\alpha') \sim 
\frac{1}{(\alpha-\alpha')^{c_{D}}}$.
A scaling argument then implies $\lambda=2$. The precise relation
between the strength $D$ of the repulsive potential and the loop
exponent $c_{D}$  depends on the behaviour of the density of states
of the Hamiltonian $\beta h_{D}$ at low energy. It can be obtained
numerically ; in particular $D$
can be chosen so that one has $c_{D}=c$ (e.g. $c=1.8$ is recovered with 
$\frac{D}{a^2} \simeq \frac{1}{3}$).

On the other hand, the orientation constraint can be written as
\begin{equation}
\label{magnetic}
\Delta
\theta=\theta_{\alpha+1}-\theta_{\alpha^{\prime}}=\int_{\alpha'}^{\alpha+1} 
\frac{x \dot y-y\dot 
x}{x^2+y^2} \ ds
\end{equation}
where $(x,y)$ are the coordinates $\vec
{\rho}$ perpendicular to the stretching force $F$. This type of
constraint arises in entangled polymers \cite{Vilgis}. 

Following a calculation of Wiegel \cite{Wiegel}, the
use of a directed 
approximation for the stretched strands enables us to write the number of
conformations $\mathcal{N}\bigl(\alpha ^{\prime },\theta _{\alpha ^{\prime
}};\alpha +1,\theta _{\alpha +1}\bigr)$ as (compare with eq. (\ref{asymp})) 
\begin{equation}
\mathcal{N}\bigl(\alpha ^{\prime },\theta _{\alpha ^{\prime }};\alpha
+1,\theta _{\alpha +1}\bigr)=\frac{\mu (F)^{\ \alpha -\alpha ^{\prime }}}{%
(\alpha -\alpha ^{\prime })^{c}}\ h(\theta _{\alpha +1}-\theta _{\alpha
^{\prime }})  \label{entropy}
\end{equation}%
where $\mu (F)=\mu _{0}\ e^{\frac{\beta ^{2}F^{2}a^{2}}{12}}$ and $h(\theta
_{\alpha +1}-\theta _{\alpha ^{\prime }})$ is a (normalized) measure of the
torsional entropy reduction given by 
\begin{equation}
h(\theta _{\alpha +1}-\theta _{\alpha ^{\prime }})=\frac{1}{\sqrt{2\pi A}}%
\exp \left( -\frac{\left( \theta _{\alpha +1}-\theta _{\alpha ^{\prime
}}\right) ^{2}}{2A}\right)   \label{torsion}
\end{equation}%
with 
$
A=\sqrt{\frac{a^{2}}{3D}}\log (\frac{4a^{2}}{3d^{2}}(\alpha -\alpha ^{\prime
}))
$.
In the previous equation, $d$ denotes the diameter of the double
helix ($d\approx 20~\mathring{A}$) and as seen above, $a$ denotes
the Kuhn length of (ss) DNA ($a \approx
50~\mathring{A}$). The experimental value of $\ \log \mu _{0}$ is
taken as $12.5$ (see ref.\cite{WB85}). For forces $F$ of order $1$ pN,
the difference between $\log \mu $ and $\log \mu _{0}$ is of order
$0.1$ and will thus be neglected in the following. The validity of the
directed approximation for this calculation will be discussed thoroughly in
a forthcoming paper \cite{tobe}.

\bigskip

In equation (\ref{z1}), the torsional energy $E_{1}\bigl(\alpha ^{\prime
},\theta _{\alpha ^{\prime 
}};\alpha +1,\theta _{\alpha +1}\bigr)$ of a bubble $(\alpha ^{\prime
},\alpha +1)$ is assumed to be of the form

\begin{equation}
E_{1}\bigl(\alpha ^{\prime },\theta _{\alpha ^{\prime }};\alpha +1,\theta
_{\alpha +1}\bigr)=\frac{\kappa _{l}}{2}\frac{\left( \theta _{\alpha
+1}-\theta _{\alpha ^{\prime }}\right) ^{2}}{\left( \alpha -\alpha ^{\prime
}\right) }
\end{equation}%
where $\kappa _{l}$ is the torsional constant of a DNA bubble.

For long enough loops ($\alpha-\alpha' >> \log (\alpha-\alpha')$),
this energy is small compared to the entropy reduction of
eq. (\ref{torsion}). Furthermore, due to the softness of unbound
fragments, one expects that $\kappa_l << \kappa_0$. In the
following, we will thus set this torsional energy to zero.

The recursion relation for the partition function therefore reads 
\[
Z(\alpha +1,\theta _{\alpha +1})=\ e^{-\beta \varepsilon _{\alpha ,\alpha
+1}}\int_{-\infty }^{+\infty }{\frac{d\theta _{\alpha }}{M_{0}}}e^{-{\frac{%
\beta \kappa _{0}}{2}}(\theta _{\alpha +1}-\theta _{\alpha }-\theta
_{0})^{2}}\ Z(\alpha ,\theta _{\alpha }) 
\]%
\begin{equation}
+\sigma _{S}\sum_{\alpha ^{\prime }=1}^{\alpha -1}\int_{-\infty }^{+\infty }%
\frac{d\theta _{\alpha ^{\prime }}}{M_{1}}\mathcal{N}\bigl(\alpha ^{\prime
},\theta _{\alpha ^{\prime }};\alpha +1,\theta _{\alpha +1}\bigr)Z(\alpha
^{\prime },\theta _{\alpha ^{\prime }})  \label{w1}
\end{equation}%
where $\mathcal{N}\bigl(\alpha ^{\prime },\theta _{\alpha ^{\prime }};\alpha
+1,\theta _{\alpha +1}\bigr)$ is given in eqs. (\ref{entropy}, \ref{torsion}%
).

Since the integration over the angular variables $\theta_{\alpha}$ should
yield back equation (\ref{y1}), the normalization factors are $M_0=\sqrt{%
\frac{2\pi }{\beta \kappa _{0}}}$ and $M_1=1$.

Setting $\theta _{1}=0$, the boundary conditions pertaining to equation (\ref%
{w1}) are
$ Z(1,\theta _{1}) =\delta (\theta _{1}) $ and
$Z(2,\theta _{2}) =\frac{1}{M_{0}}e^{-\beta \varepsilon _{1,2}-{\frac{%
\beta \kappa _{0}}{2}}(\theta _{2}-\theta _{0})^{2}}$

Equation (\ref{w1}) can be brought to the form of a standard PS recursion
relation by going to the torque representation. We define the Laplace
transform

\begin{equation}
Z(\alpha ,\Gamma )=\int_{-\infty }^{+\infty }d\theta _{\alpha }e^{\beta
\Gamma \theta _{\alpha }}Z(\alpha ,\theta _{\alpha })
\end{equation}

The quantity $Z(\alpha ,\Gamma )$ represents the partition function of a DNA
chain of length $\alpha $ fixed at the origin, and subject to a torque $%
\Gamma $. Taking the Laplace transform of (\ref{w1}), we obtain

\begin{equation}
Z(\alpha+1,\Gamma)=\ e^{-\beta \varepsilon _{\alpha,\alpha+1 }^{\prime }}\
Z(\alpha,\Gamma )+\sigma _{S}^{\prime }\sum_{\alpha ^{\prime }=1}^{\alpha -1}%
\frac{\mu_0^{~\alpha -\alpha ^{\prime }}}{\left( \alpha -\alpha ^{\prime
}\right) ^{c^{\prime }}}~Z(\alpha ^{\prime },\Gamma )  \label{PS}
\end{equation}%
with\newline

\begin{equation}
\varepsilon _{\alpha ,\alpha+1 }^{\prime }=\varepsilon _{\alpha,\alpha
+1}-\Gamma \theta _{0}-\frac{\Gamma ^{2}}{2\kappa _{0}}  \label{stack}
\end{equation}

\begin{equation}
c^{\prime }=c-\frac{\beta ^{2}\Gamma ^{2}}{2}\sqrt{\frac{a^{2}}{3D}}
\label{cp}
\end{equation}

\begin{equation}
\sigma _{S}^{\prime }=\sigma _{S}\ \left( \frac{4a^{2}}{3d^{2}}\right) ^{%
\frac{\beta ^{2}\Gamma ^{2}}{2}\sqrt{\frac{a^{2}}{3D}}}  \label{sig}
\end{equation}

The boundary conditions translate into
$
Z(1,\Gamma ) =1$ and 
$Z(2,\Gamma ) =e^{-\beta \varepsilon _{1,2}^{\prime }}$

In the form (\ref{PS}), we recognize a standard PS recursion equation, with
stacking energies given by (\ref{stack}), loop exponent given by (\ref{cp})
and loop formation (cooperativity) parameter given by (\ref{sig}). These new
effective parameters have the following properties: i) the loop exponent is
decreased by the torque, so that the probability of return to the origin is
increased (see eq.(\ref{asymp})), ii) the loop formation parameter is
increased by the torque. With realistic values of the parameters (see below), 
this effect is very weak.

These results can easily be understood for undercoiling ($\Gamma <
0$). However, in the case of strong positive supercoiling ($\Gamma >0$), the
spatial arrangements of the bases are important, a
feature which is absent of the PS approach. This can lead to the
appearance of new phases such as P-DNA \cite{Strick}.  We thus can trust our
approach for negative and weakly positive torques.

If one takes all stacking energies equal to $-\varepsilon _{0}$ , equation (%
\ref{PS}) can be solved analytically by using the appropriate grand
canonical partition function $Q(z,\Gamma )=\sum_{\alpha =1}^{\infty
}z^{\alpha }Z(\alpha ,\Gamma )$ with the results $Q(z,\Gamma )=\frac{z}{%
1-ze^{\beta \varepsilon _{0}^{\prime }}-\sigma _{s}^{\prime
}z~P{}_{c^{\prime }}(z\mu_0 )}~$where $P_{c'}(z)=\sum_{l=1}^{\infty }\frac{z^{l}%
}{l^{c'}}$. As in the original PS model, \ the critical point is obtained
when $z$ is a pole of the denominator and $z\mu_0 =1$. As
previously mentioned, the specific heat exponent (in the thermodynamic
limit) reads $\alpha (\Gamma)=\frac{2c'-3}{c'-1}$.

A theoretical outcome of our model is that the denaturation transition
disappears at large enough torque. Indeed, as we perviously saw, 
the PS model displays a phase transition only if the loop exponent $%
c$ is larger than $1$. Equation (\ref{cp}) shows that even though the bare
exponent $c$ is larger than $1$, it becomes smaller than $1$ when the torque
is increased. Therefore, the transition gets smoothed out as the torque is
increased: The denaturation peaks broaden and are shifted to lower or higher
temperatures, depending on the sign of the applied torque.
%
%



For non homogeneous stacking energies, the properties of $Q(z,\Gamma)$ 
are not amenable to analytic calculations and one has to resort to
numerical calculations. The parameters we use are $c=1.8$
(corresponding to $D/a^2 \approx 0.33$), $\sigma_S =
1.26 \ 10^{-5}$, and the MELTSIM stacking energies \cite{Meltsim}.

In Figure 3, we plot the derivative with respect to the temperature of the
fraction of bound pairs $c=-\frac{d\theta}{dT}$
(related to the experimental UV absorption of DNA) 
for a biological sequence \cite{ftp} of 2000 base
pairs as a function of temperature, for various values of the 
torque $\Gamma$. As $\Gamma$ increases, the peaks get smoothed out.

\begin{figure}[htbp]
\begin{center}
\includegraphics[scale=0.6]{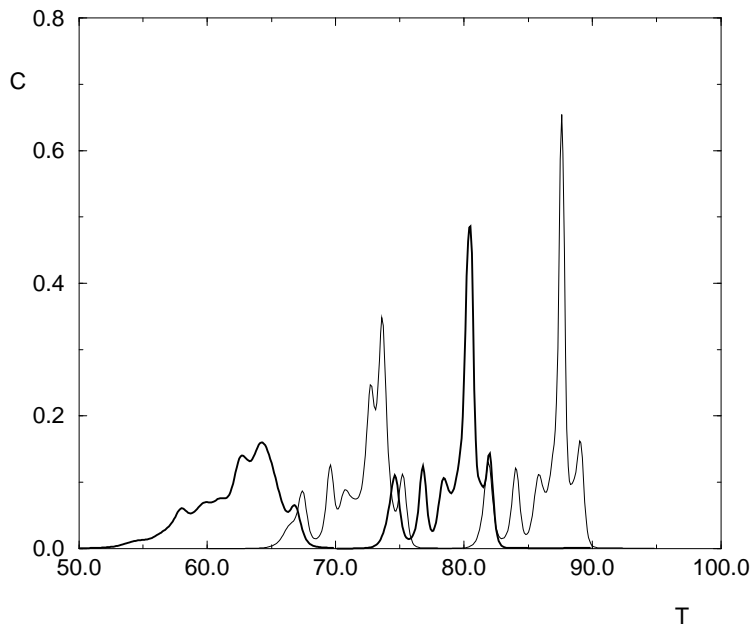}
\end{center}
\caption{Derivative of the number of paired bases $c=-\frac{d \theta(T,\Gamma)}{dT}$ 
for a 2000 bps biological fragment, for
$\Gamma=0 \ (thin), -150 \ (thick), -300 \ (thin) ,-500 \ (thick)$ K,
from right to left.}   
\label{f4}
\end{figure}


Up to here, we have considered the problem of a DNA fragment subject
to a weak torque. We now come back to the case where the DNA fragment is
supercoiled or undercoiled (as is the case of circular DNA in plasmids): 
the winding angle $\theta _{N}-\theta _{1}$ is not equal to the total
natural twist angle $(N-1)\theta _{0}$ . The supercoiling
index $s$ is defined by

\begin{equation}
\theta _{N}-\theta _{1}=(N-1)~\theta _{0}~(1+s)
\end{equation}%
It can be positive (supercoiling, $\Gamma >0$) or negative
(undercoiling, $\Gamma <0$).

The partition function $Z(N,\theta _{N})$ is related to the torque
representation through an inverse Laplace transform

\begin{equation}
Z(N,\theta _{N})=\beta \int_{C_{0}-i\infty }^{C_{0}+i\infty }\frac{d\Gamma }{%
2i\pi }e^{-\beta \Gamma (N-1)\theta _{0}(1+s)}~Z(N,\Gamma )
\end{equation}%
where $C_{0}$ is a constant which leaves all the singularities of $\
Z(N,\Gamma )$ to its right. For large $N$, one may perform a saddle
point calculation, with $\Gamma_s$ defined by 
\begin{equation}
\label{gammasad}
\frac{1}{N}\bigl(\frac{\partial \log Z(N,\Gamma )}{\partial
\Gamma}\bigr)_{\Gamma=\Gamma_s}=\beta \theta_{0}(1+s)
\end{equation}

\bigskip
For homogeneous sequences (all stacking energies equal to
$-\varepsilon _{0}$), one may go one step further since

\begin{equation}
Z(N,\theta _{N})=\beta \doint \frac{dz}{2i\pi z^{N+1}}\int_{C_{0}-i\infty
}^{C_{0}+i\infty }\frac{d\Gamma }{2i\pi }e^{-\beta \Gamma (N-1)\theta
_{0}(1+s)}~Q(z,\Gamma )  \label{t1}
\end{equation}%
where the $z~$\ integral is to be performed on a circle containing the point 
$0$.

In the limit of large $N$, equation (\ref{t1}) can again be evaluated
by the saddle point method on both $z$ and $\Gamma $ . The
saddle-point equation for $z$ is given by $N=z_{s}\frac{\partial
}{\partial z_{s}}\log Q$ . In the thermodynamic limit, $N\rightarrow
\infty $, the saddle-point solution $z_{s} $ should approach the
($\Gamma$-dependent) pole
$z^{\ast }$ of $Q(z, \Gamma)$, defined by 

\begin{equation}
1-z^{\ast }e^{\beta \varepsilon _{0}^{\prime }}-\sigma _{s}^{\prime }z^{\ast
}~P_{c^{\prime }}(z^{\ast }\mu_0 )=0  \label{pole}
\end{equation}

The saddle point (\ref{gammasad}) then reads

\begin{equation}
\beta \theta _{0}(1+s)=\bigl(\frac{\partial \log z^{\ast }}{\partial \Gamma }\bigr)_{\Gamma=\Gamma_s}
\end{equation}
and to leading  order, the free energy per base pair $f(s)$ is given by

\begin{equation}
\beta f(s)=\log z^{\ast }(\Gamma_s)+\beta \Gamma_s \theta
_{0}(1+s)
\label{f22}
\end{equation}

To make contact with previous work, we restrict ourselves to small
supercoiling ($s<<1$) or equivalently small couple $\Gamma $. Using
equation (\ref{pole}) we can expand $z^{\ast }$ to second order in
$\Gamma $ . The details of the calculations will be presented in
\cite{tobe} and we just quote the results: 

\begin{equation}
\label{benham}
\beta f(s)=\log z^{\ast}(\Gamma=0)+\frac{\beta ^{2}\theta _{0}^{2}}{%
2\Delta _{2}}\left( s+\frac{N_{u}}{N}\right) ^{2}  
\end{equation}%
where $\Delta _{2}$ is a constant, equal to the fluctuation of the end
angle ($\Delta _{2}=\bigl(\frac{\partial ^{2}\log
z^{\ast }}{\partial \Gamma ^{2}}\bigr)_{\Gamma =0}=\frac{\beta^2}{N}\bigl(
\left\langle \theta _{N}^{2}\right\rangle -\left\langle \theta
_{N}\right\rangle ^{2}\bigr) _{\Gamma =0}$)

The physical interpretation of equation (\ref{benham}) is fairly simple. It
essentially states that for small supercoiling $s$, the free energy of a DNA
strand is equal to the sum of the free energy of the non supercoiled
fragment plus a free energy term which forces the fraction of unbound pairs
to be close to the opposite of the supercoiling index. This form is very
similar to the form which was devised phenomenologically by Benham \cite%
{Benham}. In our case, this form is derived from the microscopic model, and
the coefficients of the quadratic part are expressed in terms of the
microscopic characteristics of the DNA.

For larger supercoiling, the relation between $s$ and the fraction of
unbound pairs is not straightforward. Arguments that suggest that $-s \sim
\frac{N_u}{N}$  for any $s$ are given in ref. \cite{Bar_Le_Pey_Theo}.

Finally, we point out that the specific heats $C(T,\Gamma)$ and
$C(T,s)$ are related through Fisher renormalization
\cite{Fisherrenorm}. In particular, if $C(T,\Gamma)$ diverges,
$C(T,s)$ will be finite at the transition.

\begin{figure}[htbp]
\begin{center}
\includegraphics[scale=.7]{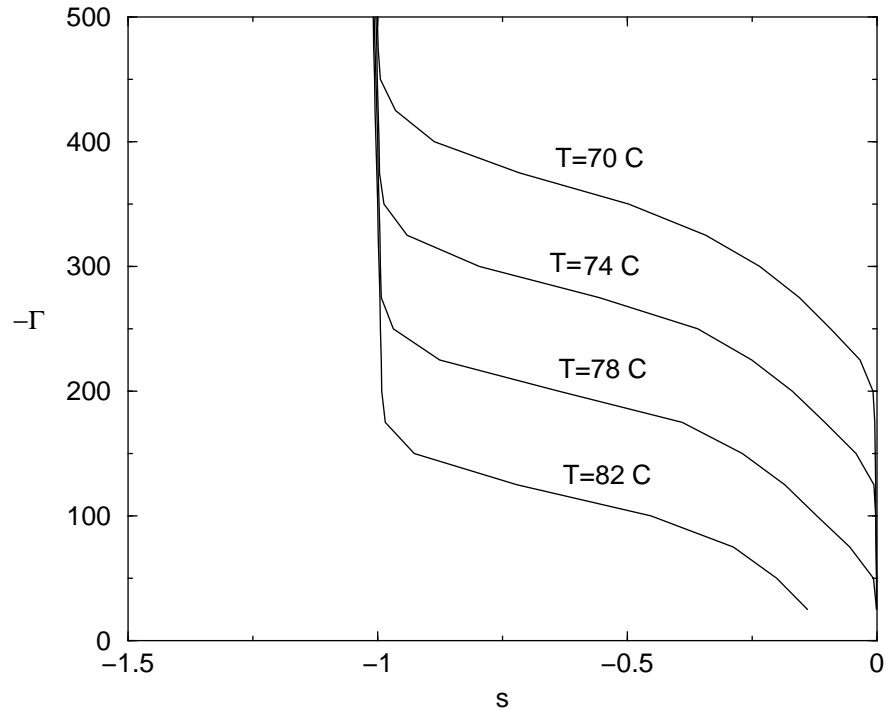}
\end{center}
\caption{Torque as a function of the supercoiling index for various
  temperatures, for the sequence of Figure 3.}
\label{f5}
\end{figure}

For non homogeneous sequences, numerical calculations yield
$Z(N,\Gamma)$, and through eq.(\ref{gammasad}), one obtains the
isotherm curves $\Gamma(s)$. The application to the sequence used in
Figure \ref{f4} is shown in Figure \ref{f5}.

The statistical mechanics of the torque induced DNA denaturation has
been previously studied in different contexts
\cite{Co_Mo,Ru_Bru,Hwa,Strick}, and there is agreement with our results
whenever they overlap. Ref \cite{Co_Mo} considers a transfer matrix
formalism for homogeneous sequence with $c=\frac{3}{2}$.
Ref \cite{Ru_Bru} starts from the Benham phenomenological expression
and obtains - among other results - the critical curve $T_c(s)$ in the
homogeneous case. Ref \cite{Hwa} studies the thermodynamic and dynamic
properties of a random DNA fragment submitted to a torque.
Finally Ref \cite{Strick} discusses the experimental procedure in detail
(ranges of force and torque); the comparison with our results is not
straightforward as  experiments rely on the extension curve, which is
not included in PS type of models.

We have proposed a generalization of the PS model to include the
effect of torsion on DNA denaturation. Torsion 
produces a very strong reduction of entropy in the loops, which
eventually suppresses the denaturation transition.
Application to cyclic DNA (plasmids) will be the subject of future work.


\begin{thebibliography}{99}

\bibitem{CW} J.D. Watson and F.H.C. Crick, Nature, \textbf{171}, 737 (1953).

\bibitem{Pol_Scher2} D. Poland and H.A. Scheraga eds., \textit{Theory of
Helix-Coil transition in Biopolymers}, Academic Press, New York (1970).

\bibitem{Pol_Scher1} D. Poland and H.A. Scheraga, J. Chem. Phys., \textbf{45}%
, 1456, 1464 (1966).

\bibitem{Meltsim} R.D. Blake, J.W. Bizarro, J.D. Blake, G.R. Day, S.G.
Delcourt, J. Knowles, K.A. Marx and J. SantaLucia Jr., Bioinformatics, 
\textbf{15}, 370-375 (1999).

\bibitem{Strick} T.R. Strick, M-N. Dessinges, G. Charvin, N.H. Dekker,
J-F. Allemand, D. Bensimon and V. Croquette, Rep. Prog. Phys.,
\textbf{66}, 1 (2003).  

\bibitem{PGG} P.G. de Gennes, \textit{Scaling concepts in polymer physics} ,
Cornell University Press, Ithaca, (1979).

\bibitem{Fisher} M.E. Fisher, J. Chem. Phys., \textbf{45}, 1469 (1966).

\bibitem{Ka_Mu_Pe} Y. Kafri, D. Mukamel and L. Peliti, Phys. Rev. Lett., 
\textbf{85}, 4988 (2000).

\bibitem{Dup} B. Duplantier,  Phys. Rev. Lett., {\bf 57}, 941 (1986);
J. Stat. Phys., {\bf 54}, 581 (1989).

\bibitem{Orlandini} E. Carlon, E. Orlandini and A.L. Stella,
Phys. Rev. Lett., {\bf 88}, 198101 (2002).

\bibitem{Poland} D. Poland, Biopolymers, \textbf{13}, 1859 (1974).

\bibitem{Ga_Orl} T. Garel and H. Orland, q-bio.BM/0402037.

\bibitem{tobe} T. Garel, H. Orland and E. Yeramian, to be published.

\bibitem{Vilgis} A.L. Kholodenko and T.A. Vilgis, Phys. Repts., {\bf
298}, 251 (1998) and references therein.

\bibitem{Wiegel} F.W. Wiegel, in \textit{Phase Transitions and Critical
Phenomena}, C. Domb and G. Green eds, Vol. 7, p. 101, Academic Press (1983)
and references therein.

\bibitem{WB85} R.M. Wartell and A.S. Benight, Phys. Repts.,
\textbf{126}, 67 (1985).

\bibitem{ftp} This sequence consists of lines 34-67 of AE014135
(Drosophila melanogaster chromosome 4) at
http://www.ebi.ac.uk/genomes/eukaryota.html.

\bibitem{Benham} C.J. Benham, Phys. Rev. E, \textbf{53}, 2984 (1996) and
references therein.

\bibitem{Bar_Le_Pey_Theo} M. Barbi, S. Lepri, M. Peyrard and
N. Theodorakopoukos, cond-mat/0309454.

\bibitem{Fisherrenorm} M.E. Fisher, Phys. Rev., {\bf 176}, 257 (1968).

\bibitem{Co_Mo} S. Cocco and R. Monasson, Phys. Rev. Lett., \textbf{83},
5178 (1999).

\bibitem{Ru_Bru} J. Rudnick and R. Bruinsma, Phys. Rev. E, \textbf{65},
(R)030902 (2002).

\bibitem{Hwa} T. Hwa, E. Marinari, K. Sneppen and L-h. Tang, Proc. Natl.
Acad. Sci. (USA), \textbf{100}, 4411 (2003).



\end{thebibliography}
\end{document}